\documentclass[sigconf, nonacm, screen]{acmart}

\usepackage{include/packages}
\usepackage{include/commands}
\usepackage{include/shorthands}

\begin{document}

\pdfpagewidth=8.5in
\pdfpageheight=11in

\pagenumbering{arabic}

\title{NUNA: Characterizing and Mitigating Non-Uniform Network Access in Multi-Die GPU Scale-Up Systems
\vspace{0.25em}
}

\author{%
Conor~James~Green$^{1,2}$,\space{}
William~Won$^{1}$,\space{}
Tuan~Ta$^{1}$,\space{}
and~Bradford~M.~Beckmann$^{1}$
}
\affiliation{%
\vspace{0.5em}
\institution{%
    $^{1}$AMD Research and Advanced Development\space{}\space{}
    $^{2}$Purdue University
}
\city{}
\country{}
\vspace{1em}
}
\thanks{%
    Correspondence to:
    Conor Green <Conor.Green@amd.com, green456@purdue.edu>,
    William Won <William.Won@amd.com>.
}

\begin{abstract}

Graphics processing unit~(GPU) architectures are growing in size to meet the increasing compute and memory requirements.
As GPU sizes increase, intra-socket wire transfer delay increases significantly.
While previous research has optimized for compute and memory locality within a socket, the spatial impact on inter-GPU communication has not been well-studied.
We introduce the term non-uniform network access~(\ours{}) to describe this emerging optimization dimension in multi-GPU systems.
We specifically focus on latency-sensitive collective communication, common in machine learning inference.
First, we highlight the need for \ours{}-aware routing~(\nar{}), choosing optimized, spatially-aware inter-GPU paths in large scale-up network topologies.
Second, we introduce \ours{}-aware placement~(\nap{}), placing threadblocks and data near I/O to optimize the inter-GPU traffic.
We demonstrate that the \nap{} optimizations alone offer up to 1.5$\times$ collective speedups over a locality-unaware baseline.
Combining \nap{} with \nar{} yields up to 1.8$\times$ faster collectives over the locality-unaware baseline.
This leads to 7\%~mean (28\%~max) time~per~output~token speedup in machine learning inference.
\end{abstract}

\maketitle
\thispagestyle{plain}
\pagestyle{plain}

\section{Introduction}\label{sec:intro}

Artificial intelligence~(AI) models have been growing in size and require significantly more compute and memory than what a single graphics processing unit~(GPU) can provide.
Prominent large language models~(LLMs) such as GPT-4~\cite{openai2024gpt4technicalreport} and Llama-4~\cite{meta2025llama} have billions to trillions of parameters, requiring significant on-chip memory capacity and compute   capability.
To meet this thrust for AI compute and memory throughput, designers have recently adopted aggressive packaging technology where individual GPUs are composed of multi-chip modules~(MCMs) and 3D-integrated chiplet architectures~\cite{nvidia2025gtc,amd2025advancingai}.
With the AI chip market expected to expand to hundreds of billions of annual sales by the 2030s~\cite{correa2025mlmarket,martina2025analystday}, hardware vendors can affordably leverage more extravagant power delivery and cooling solutions, integrating even more reticle-limit compute and memory dies into a single package~\cite{patel2023ai,smith202411,smith2024amd}.
The resulting multi-die GPUs provide significant improvements by doing more work per socket.
However, the longer physical distances between compute units~(CUs), memory, and input/output~(I/O) ports significantly impact intra-socket latency~\cite{jin2024uncovering,luo2024benchmarking}.
While recent work has highlighted optimizations for intra-socket non-uniform memory access~(\numa{}) locality~\cite{tee2025mall,amdDeepDive}, the impact of intra-socket latency on inter-GPU communication has not been well-studied.

To further satisfy the massive AI compute and memory demands, hardware vendors have recently introduced novel scale-up networks where multiple GPUs can communicate across a single shared address space using load and store instructions~\cite{geeyarpuram2025nvlinkfusion_slides}.
This enables AI models to be distributed across multiple GPUs through parallelization techniques~\cite{openai2024gpt4technicalreport, zhang2022opt, meta2025llama, brown2020language}.
However, the parallelization requires collective communications among GPUs to communicate activations and gradients, which can consume more than 40\% of total execution time~\cite{hwang2023tutel,pati2024t3,mudigere2023SoftwarehardwareCodesignFast}.
Hardware vendors continue to improve direct GPU-to-GPU interconnects, such as AMD Infinity Fabric\texttrademark{}~\cite{schieffer2024understanding} or NVIDIA NVLink/NVSwitch technologies~\cite{li2019evaluating}, to address this scale-up network bottleneck.
Specifically, scale-up network bandwidth has increased across GPU generations by increasing the number of I/O ports per socket and data rates per port.
Network traffic is hashed across the socket's I/O ports for load balancing~\cite{nvidia2016p100,amd2021cdna,ROCmMI250docs,nvidiasgxls10,nvidia2022nvlink,nvidia2025jax,milic2017beyond}.

\insertFigure{NearvsFar_TwoGPUs}{fig:nuna_mcm_twogpu_analytical}{0.93}{-0.2em}{-1.3em}{
Possible high and low latency inter-GPU communication paths.
A threadblock~(blue) is copying from a local source~(orange) to remote destination~(blue) buffers.
Scenario~\textcircledblack{a} represents physically far accesses between CU, memory, and I/O ports, whereas \textcircledblack{b} shows low-latency accesses.
}

Overall, while the proliferation of large multi-die GPU sockets connected together using massive scale-up networks is an unprecedented, exciting time in computer architecture, their non-uniform communication latencies must be addressed to reach the full capability of these impressive systems.
In the research community, NUMA and non-uniform cache access~(\nuca{}) effects within a single socket have been analyzed for decades~\cite{luo2024benchmarking,hardavellas2009reactive,choquette2022nvidia,beckmann2004managing,kim2002adaptive} and have recently been extended to distributed systems~\cite{schieffer2024understanding,li2019evaluating,arunkumar2017mcm,fatima2025netcrafter}.
These prior NUMA and NUCA solutions have mainly focused on reducing the initial and subsequent data access penalties within a socket or a small handful of sockets. NUMA or NUCA solutions have not looked at collective communication and the multiple communication paths provided by large scale-up networks.
Specifcially, to the best of our knowledge, prior NUMA works that focused on GPUs~\cite{cabezas2015automatic,chen2017improving,arunkumar2017mcm,kim2017coda,milic2017beyond,khairy2020locality,zhao2023nuba,zhu2024spgpu,fatima2025netcrafter} have not considered the physical location of the I/O ports.

This paper is the first to address the \emph{non-uniform network access~(\ours{})} latencies arising in large multi-die GPUs and their impact on scale-up communication.
In the same way that NUMA optimizations spurred decades of research in high-performance computer architectures, we believe \ours{} will be a critical element for optimizing future high performance multi-GPU systems.

\autoref{fig:nuna_mcm_twogpu_analytical} illustrates the impact spatial distances have on inter-GPU communication.
For a threadblock to \texttt{put} data off-chip, the far paths (highlighted in red) traverse significantly more on-chip interconnect than the near paths (highlighted in green).
Imagine that collectives are na\"{i}vely implemented without considering \ours{} effects.
The collective begins with (\textcircledblack{1a})~a GPU threadblock~\texttt{a} reading a value from the local high bandwidth memory~(HBM) stack located physically far away.
Once the value is read, the value is then written remotely by (\textcircledblack{2a})~first routing it to a far I/O port, statically chosen based on address, and then (\textcircledblack{3a})~traversing the scale-up network and storing the value in the remote GPU's HBM stack located far from the I/O port it was received on.
In contrast, the \ours{}-aware collective begins with (\textcircledblack{1b})~a threadblock~\texttt{b} reading the local value from the HBM stack closest to it.
Then, (\textcircledblack{2b})~the threadblock writes the data remotely by communicating across the I/O port closest to it and (\textcircledblack{3b})~storing the data to the HBM stack closest to the receiving I/O port.
~\autoref{sec:nuna-impact} evaluates \ours{}'s impact on current generation GPUs and identifies the best and worst case communication latencies vary by almost 2$\times$. 
Quantifying \ours{}'s impact for next-generation GPUs shows that the worst- and average-case network-on-chip~(NoC) latencies can differ by 3$\times$ and 2$\times$, respectively, compared to the best case.
This results in uncontended GPU-to-GPU transfer %
slowdowns of 1.8$\times$ and 1.45$\times$ for the worst and average cases, respectively.

To systematically exploit these non-uniform intra-socket latencies for inter-GPU communications, we introduce two policies that ensure collectives leverage the best possible communication paths.
First, we introduce \emph{\ours{}-aware routing}~(\nar{}) to prioritize latency-sensitive traffic, utilizing physically closer I/O ports.
We find that applying \nar{} alone can improve collective performance by up to 35\%.
Second, we introduce \emph{\ours{}-aware placement}~(\nap{}) of threadblocks and memory pages.
\nap{} complements \nar{} by mapping threadblocks and data to CUs and HBM stacks physically closer to the I/O ports.
\nap{} and \nar{} applied together can reduce collective execution time by up to 80\%.

To summarize, we make the following contributions:
\begin{itemize}
\item Define and analyze \ours{} effects for scale-up systems of physically large GPUs.
\item Develop a \ours{}-aware routing policy, \nar{}, that requires minimal hardware changes and significantly reduces intra-GPU I/O port access latency for inter-GPU traffic optimization.
\item Propose a \ours{}-aware placement policy, \nap{}, which places threadblocks and memory pages closer to the I/O ports to reduce inter-GPU communication latency.
\item Evaluate \nar{} and \nap{} to accelerate state-of-the-art collective communication.
\end{itemize}

\section{Background}\label{sec:background}
\subsection{Collective Communications}\label{sec:collective_communication}

\insertFigure{all_both_traces_collectives_combined_cdf}{fig:cdf_traces}{0.95}{-.2em}{-1em}{
Cumulative distribution of collective sizes (in output buffer size) for inference and training workloads.
}

Collective communications have existed in high-performance computing~(HPC) for decades~\cite{mckinley1995collective,banikazemi1998efficient} and have recently been adopted by distributed AI training and inference.
Both the AI model and the parallelization techniques used to distribute the model determine the required collectives.
For example, mixture-of-experts~(MoE) models use \alltoall{} to route tokens to experts and collect results~\cite{hwang2023tutel}.
Meanwhile, data and tensor parallelization require (\rom{1})~\allreduce{} for gradient aggregation and (\rom{2})~\allgather{} to assemble split model weights~\cite{ott2021fsdp}.
Therefore, although there exist additional collective patterns, \allgather{}, \allreduce{}, and \alltoall{} are the most common patterns used by AI.

\subsection{Latency-Sensitive Collectives}

Analytical $\alpha$--$\beta$ modeling (using parameters in~\autoref{sec:motivation_inter}) shows that transmitting less than 100\,MB of traffic over multi-terabyte scale-up links contributes only microseconds of serialization delay.
This makes point-to-point latencies meaningfully impact the total communication time.
Therefore, \ours{}-aware collective optimizations show the most benefit on such latency-sensitive collectives, as they reduce the overall time for inter-GPU communication.

\paratitle{Inference}
Inference workloads account for the majority of the computational resources at data centers for companies such as Meta, Amazon, and Google~\cite{wu2022sustainable,barr2019amazon,patterson2022carbon}.
Unlike training, inference workloads (\rom{1})~often use small batches, and (\rom{2})~their decode operations operate token-by-token, resulting in small-size collectives.
Furthermore, for quality of service guarantees,
inference tasks need to optimize for latency such as time per output token~(TPOT)~\cite{nvidia2025jax}.
Recent LLM-serving systems, vLLM~\cite{kwon2023vllm} and SGLang~\cite{zheng2024sglang}, characterize decode as limited by per-iteration latency, rather than aggregate compute.
These factors make collectives in inference more sensitive to network latency rather than network bandwidth.
Consequently, optimizing latency-sensitive collectives is a critical performance challenge~\cite{erdil2025inference,singh2025big,agrawal2024taming,zhao2024hetegen}.
For instance, Google highlighted network latency optimization as one of the biggest hardware challenges that the AI industry
faces for inference~\cite{ma2026challenges}.

\autoref{fig:cdf_traces} analyzes collective sizes in end-to-end prefill and decode from public models with different numbers of parameters and parallelization for 64~GPUs. The methodology is explained in~\autoref{sec:method_llm} and the plots provide the cumulative distribution of output buffer sizes for each workload.
Notably, the prefill phase introduces larger collectives, but their size is limited to tens of megabytes.
The decode phase issues much smaller collectives, typically in the kilobytes range due to its token-by-token execution nature.
\ours{} significantly impacts these small, latency-bound collectives in LLM inference.

\paratitle{Training}
While inference is more latency-sensitive, training often includes small collectives as well.
Prior works have found that communication dominates end-to-end LLM training time~\cite{jia2024pccl,qin2025optimizing,dryden2018aluminum,pati2024cross}.
While there exists a wide diversity of model and parallelism approaches, prior work consistently finds latency-sensitive collectives in training~\cite{dryden2018aluminum,yang2020training,jia2024pccl,gangidi2024rdma}.
To quantify the collective sizes used in training, we analyze public Chakra traces from MLCommons~\cite{sridharan2023chakra,hawks2025mlcommons} and the Scalable Parallel Computing Lab~\cite{spcl_atlahs,shen2025atlahs}. These traces were collected from real systems spanning 2--256 GPUs and feature several different parallelization strategies. 
\autoref{fig:cdf_traces} plots the collective size distribution.
Across all traces, the majority of collectives are at or below 100\,MB and several traces (e.g., ResNet, DLRM, and Mixtral variants) are mostly below 10\,MB.
In particular, the tensor-parallel traces (\eg{} dense Llama- and GPT-class models) are dominated by larger (100\,MB) \reducescatter{} and \allgather{} collectives, while the data-parallel traces (\eg{} ResNet, DLRM) are dominated by smaller (10\,MB) \allreduce{} collectives.

\section{Motivation}\label{sec:motivation}

\subsection{Scaling Trends}\label{sec:problem_statement}

\subsubsection{Large Multi-Die GPU Socket}
\label{sec:motivation_hardware}

To meet AI training and inference demands, vendors have increased compute and memory capacity per socket by adding higher throughput CUs and more HBM.
Advanced packaging techniques enable integrating multiple dies per socket.~\cite{smith202411,wikichipChiponWaferonSubstrateCoWoS}.
The physical size of a single GPU continues to increase, whether they scale with MCMs~\cite{schieffer2024understanding,arunkumar2017mcm}, chiplet-based devices~\cite{smith202411,smith2024amd}, or wafer-scale systems~\cite{pal2019architecting}.
Current devices such as NVIDIA Hopper/Blackwell~\cite{luo2024benchmarking,choquette2022nvidia,nvidiaNVIDIAHopper,tirumala2024nvidia} and AMD Instinct\texttrademark{} MI300X/MI350X \cite{smith202411,smith2024amd,amdmi350x} GPUs are considerably larger and more power hungry than the prior generations.
Recent public announcements from vendors show that the pace of GPU scaling is only expected to increase~\cite{amd2025advancingai,nvidia2025gtc}.

Vendors first prioritize providing ample memory bandwidth.
GPUs expand along the long dimension, to place compute dies adjacent to as many HBM stacks, resulting in an elongated cigar-shaped socket with significant signaling beachfront.
Consequently, inter-GPU I/O ports are located along the far edges.
\autoref{fig:nuna_evaluation_system} depicts a representative next-generation GPU design.

\insertFigure{nuna_evaluation_system}{fig:nuna_evaluation_system}{.75}{-0.4em}{-1.0em}{
Representative scale-up system evaluated in this work.
Each GPU socket consists of four reticle-limit compute dies~(yellow) with 16 HBM stacks~(purple) along the perimeter.
The inter-GPU network is a single-level Clos.
}

\subsubsection{System Scaling Through Scale-Up Network}
\label{sec:motivation_inter}

In addition to scaling a single GPU, modern AI systems leverage compute and memory across multiple devices.
Specifically, scale-up networks create a shared memory with a single address space.
The scale-up domain features relatively higher throughput and lower latency, compared to scale-out networks.

\autoref{fig:nuna_evaluation_system} illustrates a representative scale-up system (we omit the I/O ports to the CPU host or scale-out NICs for brevity)~\cite{amd2025advancingai, nvidia2025gtc,nvidiaNVLinkNVSwitch,naddodNVIDIAGB200}.
A single-level Clos topology connects multiple sockets~\cite{clos1953study}.
Each scale-up switch connects to the I/O ports with the same ordinal ID, creating an independent communication plane~(\ie{} rail-optimized).

\subsubsection{Intra-Socket versus Scale-Up Latency}

Intra-socket latencies are growing as the physical footprint scales.
NoC latency
significantly increases as on-chip wire pitches scale down~\cite{beckmann2004managing}.
However, scale-up link and switch latencies are expected to shrink (or stay constant)~\cite{ualinkSpec,lutz2020pumpup,jung2025compute}.
Low-attenuation inter-socket transmissions are less sensitive to physical distance~\cite{turner2018ground}.
Low-latency scale-up technologies, such as optical networks, further exemplify this.

These trends made the intra-socket NoC and scale-up network latencies comparable.
We analyzed the NoC and scale-up latencies for \autoref{fig:nuna_evaluation_system},
using recent commercial multi-die GPU~\cite{jin2024uncovering,luo2024benchmarking,luhnen2024benchmarking} and scale-up network~\cite{ualinkSpec,lutz2020pumpup,jung2025compute} values.
For a remote write, the NoC transfer takes up to 0.9\,$\mu$s, while the scale-up latency is 
1\,$\mu$s.
Consequently, NoC transfer delay can greatly impact the total inter-GPU communication latency, when the communication is latency-sensitive.
\begin{figure}[t]
    \centering

    \begin{tabular}{ccc}
    \centering
    \begin{subfigure}{0.28\linewidth}
        \centering
        \includegraphics[width=\linewidth]{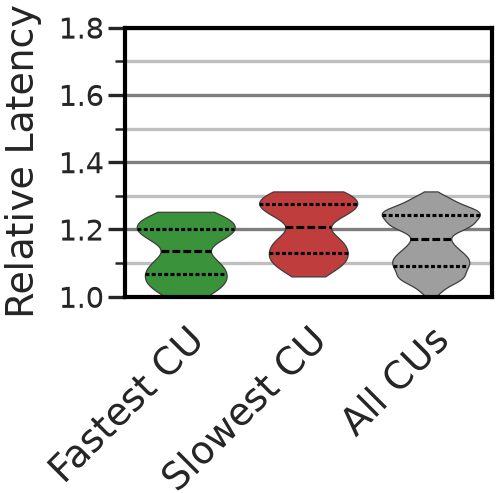}
        \vspace{-1.5em}
        \caption{AMD Instinct\texttrademark{} MI210}
        \label{fig:gpu_profiling_mi210}
    \end{subfigure}  &
    \begin{subfigure}{0.28\linewidth}
        \centering
        \includegraphics[width=\linewidth]{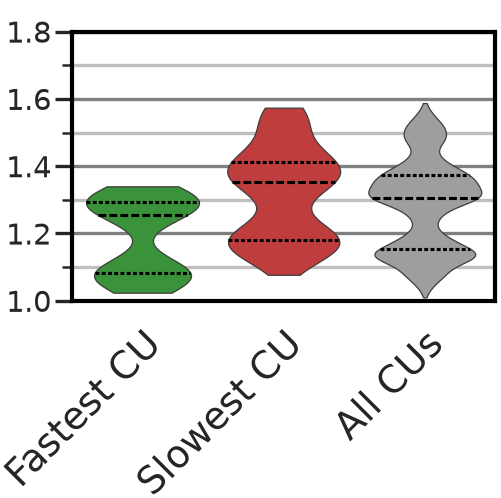}
        \vspace{-1.5em}
        \caption{AMD Instinct\texttrademark{} MI355X}
        \label{fig:gpu_profiling_mi355x}
    \end{subfigure} &
    \begin{subfigure}{0.28\linewidth}
        \centering
        \includegraphics[width=\linewidth]{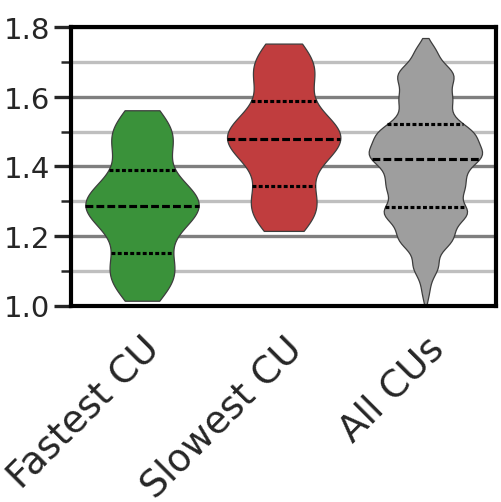}
        \vspace{-1.5em}
        \caption{AMD Instinct\texttrademark{} MI300X}
        \label{fig:gpu_profiling_mi300x}
    \end{subfigure} \\
    \end{tabular}
    \vspace{-.6em}

    \caption{
    Real-system measured remote store latencies between two GPUs, across three clusters.
    The width indicates frequency and the dashed lines indicate quartiles.
    }
    \label{fig:gpu_profiling}
    \vspace{-1.2em}
\end{figure}

\subsection{Non-Uniform Latency for Inter-GPU Communication}\label{sec:nuna-impact}

The recent scaling trends emphasize the importance of understanding intra-socket latency for inter-GPU communications.
However, 
there has been little prior evaluation on how the CU and memory location affects remote access latency.

To demonstrate intra-socket latency effect on inter-GPU communication, we profiled real-system scale-up communication latencies.
Specifically, we profiled three scale-up systems, each with different generations of GPUs: AMD Instinct\texttrademark{} MI210, MI355X, and MI300X.
Each socket has one, two, and four interposer dies, respectively.
We issued an atomic \texttt{add} operation from a CU to a remote GPU target address and measured the latency.
To profile the spatial effect, we repeated this process over all CUs and target address range of a GPU, and measured the latency variations.
The results are plotted in~\autoref{fig:gpu_profiling}.

The profiling results demonstrate that the remote communication latency significantly varies by local compute and data locations.
We observe multiple trends.
First, the communication latency variation is proportional to the physical size of the socket (\ie{} number of interposer dies).
Compared to the single-die AMD Instinct\texttrademark{} MI210, the four-die Instinct\texttrademark{} MI300X has 3$\times$ and 5$\times$ higher variation amongst target addresses and CUs, respectively.
As vendors physically scale next-generation GPUs, the non-uniformness becomes more prominent and should be addressed for latency-sensitive communications.
Second, some CUs have overall lower remote communication latency.
\autoref{fig:gpu_profiling} shows that the latency gap between the fastest and slowest CUs, especially the tail latency, increases as the socket grows.
Collectives are tail-latency-bound.
Therefore, some CUs are better than others and should be prioritized for latency-bound collectives.
Finally, the memory location also varies the remote access latency.
A single CU still experiences significant latency variation dependent on the target address.
Further reducing the tail latency requires considering physical data placement in HBMs.

To summarize, the physical placement of compute and data significantly impacts the inter-GPU transfer latency.
The spatial impact especially increases as the socket size scales.
Therefore, we must mitigate this \ours{} effect to optimize the latency-sensitive collective performance.

\subsection{Limitations of NUMA Optimizations}\label{sec:prior_numa_optimizations}

While~\numa{} mitigations colocate compute with the memory it accesses, \ours{} introduces a new dimension: I/O.
Addressing \ours{} requires colocating I/O in between compute and memory.
Thus, we purposely introduced the term \ours{} to highlight this parallel nature to \numa{}.

Concretely, prior~\numa{} techniques for distributed GPUs cannot adequately address \ours{} effects.
\autoref{tab:numa_vs_nuna} compares them.
\numa{} techniques lack (\rom{1})~hardware implementations to control off-chip routing, (\rom{2})~analysis of remote buffers, and (\rom{3})~algorithms to improve communication threadblock scheduling.
These three deficiencies correspond to our three contributions: \nar{}, \nap{} memory placement, and \nap{} threadblock scheduling.

\paratitle{Fine-Grain Routing}
Existing NUMA techniques ignore distinct I/O ports, combined with CU and memory locations, show non-uniform network latencies.
Without considering such non-uniformity across I/O ports, locality optimizations cannot mitigate \ours{}.
For example, Khairy~\etal{} address hierarchical 
applications through socket-level colocation~\cite{khairy2020locality}, and Milic~\etal{} discuss fine-grained I/O port assignment to dynamically adjust directional bandwidth.
However, neither addresses non-uniform latency across I/O ports nor the routing scheme to mitigate such effect.

\paratitle{Remote Buffer Locality}
While previous NUMA works colocate compute to local buffers, none optimize remote data placement, especially relative to I/O~\cite{cabezas2015automatic,kim2017coda,li2017locality,khairy2020locality,kim2023locality,zhu2024spgpu}.
Previous techniques do not offer methods to resolve the conflict between increasing locality (conventionally achieved by placing data near accessing threadblocks) and keeping buffers remote (placed/pinned at their respective devices).

\begin{table}[t]
\caption{Summary of \numa{} and \ours{} techniques.}
\vspace{-1em}
\label{tab:numa_vs_nuna}
\begin{tabular}{lcc}
\toprule
\textbf{Mitigation Technique}  & \textbf{NUMA}              & \textbf{NUNA}              \\ \midrule
Local buffer locality to compute         & \ding{52} & \ding{52} \\
Fine-grain off-chip routing          & \ding{56} & \ding{52} \\
Remote buffer locality to I/O        & \ding{56} & \ding{52} \\
Threadblock scheduling for remote traffic & \ding{56} & \ding{52} \\ \bottomrule
\end{tabular}
\vspace{-0.5em}
\end{table}

\paratitle{Communication Threadblock Placement}
Prior NUMA works often assume coarse-grained memory allocations to GPU sockets~\cite{kim2017coda,kim2023locality} or only focus on intra-socket locality~\cite{chen2017improving,li2017locality} without distinguishing local and remote access patterns.
They simply group threadblocks accessing the same data, and distribute them evenly across the entire GPU using simple round-robin scheduling~\cite{arunkumar2017mcm,khairy2020locality}.
However, the communication threadblocks should be allocated to the CUs physically close to the I/O ports.
This is especially true for latency-bound small collectives where each threadblock has very low resource usage but high latency sensitivity.

\section{Optimizing for \ours{}}\label{sec:optimizing_for_nuna}

Optimizing for \ours{} requires multiple steps:

\begin{itemize}[leftmargin=*]
    \item A compiler or profiling process decomposes the collective plan to identify each threadblock's loads and stores.
    \item \ours{}-aware placement (\nap{}) groups threadblocks and the chunks they access into logical groups and round-robin assigns them to \ours{}-aware routing (\nar{}) domains.
    \item Using the logical assignments, \nap{} allocates threadblocks to CU(s) and memory chunks to HBM stacks.
    \item The dispatcher and memory driver implement threadblock-to-CU(s) and chunk-to-memory-stack mappings via CU masking and physical address placement, respectively.
    \item The collective executes with \nap{}-optimized local and remote locality, utilizing \nar{}'s low-latency off-chip routing.
\end{itemize}

We first define and explain \nar{} as the foundation for \ours{} optimization.
Next, we describe the \nap{} algorithm, exploiting the express I/O port access capabilities from \nar{}.
\nap{} analyzes a collective communication plan and decides (\rom{1})~the best \nar{} granularity, (\rom{2})~threadblock placement, and (\rom{3})~chunk-to-page mappings. The algorithm decouples access patterns (logical allocation) from physical placement to achieve high locality while reducing resource oversubscription.

The complete \ours{}-aware optimization scheme only require modest, backward-compatible changes across the stack.
First, the \nap{} algorithm needs the backend communication library/compiler (\eg{} MSCCL++~\cite{shah2025mscclpp}, NCCL~\cite{nvidiaNVIDIACollective}, or RCCL~\cite{githubGitHubROCmrccl}) to provide collective plans. We rely on existing collective plans and leave automatic collective detection via static code analysis~\cite{kim2017coda,khairy2020locality,cabezas2015automatic,li2017locality,chen2017improving,kim2023locality,zhu2024spgpu} to future work.
For placement, we utilize threadblock-to-CU affinity masks to prioritize a threadblock to be located over a specific CU.
The driver controls memory page placement at HBM stack granularity.
Both mechanisms build on prior work that coordinates fine-grained threadblock-to-CU and virtual-to-physical mappings~\cite{vijaykumar2018locality,khairy2020locality,li2017locality,coppock2025lithos}.
For the network, we propose modifying the address hashing logic to to limit the inter-GPU communication to a subset of I/O ports.
Critically, all proposed modifications can run simultaneously with other workloads and be disabled for non-latency-sensitive applications.
\autoref{sec:nap_implementation} provides further implementation details.

\insertFigure{nuna_phys_addr_nars_v2}{fig:phys_addrs_and_nar}{0.8}{-.4em}{-1.0em}{
Static partitioning some of the physical address space into segments of various \nar{} granularities, leaving the rest unmodified~(baseline). Each \nar{} domain~(color) hashes into a subset of I/O ports.
}

\subsection{\ours{}-Aware Routing}\label{sec:nuna_aware_routing}
We propose~\nar{} to directly address the unnecessarily long paths, and NoC congestion thereby, of inter-GPU flits between CUs and I/O ports.
The baseline routing algorithm distributes off-chip traffic across all I/O ports for load balancing.
Instead, \nar{} routes flits only to a subset of I/O ports physically closer to the CU to minimize the internal latency.

\subsubsection{Baseline I/O Routing Policy}
For pod-scale systems, a single-level Clos is emerging as the dominant inter-GPU topology since it achieves one-hop routing for peer-to-peer communication.
Each I/O port per GPU attaches to a separate switch plane~\cite{nvidiaNVIDIAHopper,naddodNVIDIAGB200}. The address to I/O ports mapping is typically determined by a boot-time manager (e.g., NVIDIA Fabric Manager~\cite{nvidiaNVIDIAFabric}).
Commonly, it statically hashes lower-order physical address bits.
This simple approach balances scale-up loads across all available ports, achieving improved bandwidth.
\autoref{fig:phys_addrs_and_nar} illustrates an example of this baseline hashing scheme, where the white segment uniformly hashes physical addresses to all (four) possible I/O ports.

\begin{figure}[t]
    \centering

    \begin{tabular}{cc}
    \begin{subfigure}{0.3\linewidth}
        \centering
        \includegraphics[width=.85\linewidth]{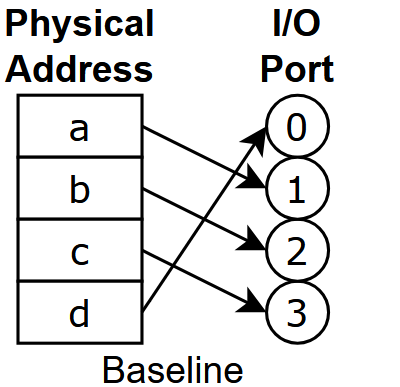}
        \vspace{-.5em}
        \caption{}
        \label{fig:nar_diagram_baseline_hashing}
    \end{subfigure} &
    \begin{subfigure}{0.3\linewidth}
        \centering
        \includegraphics[width=.85\linewidth]{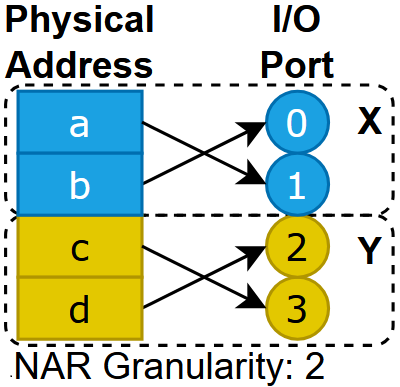}
        \vspace{-.5em}
        \caption{}
        \label{fig:nar_diagram_nar_hashing}
    \end{subfigure}
\\

    \end{tabular}

    \begin{subfigure}{\linewidth}
        \centering
        \includegraphics[width=0.85\linewidth]{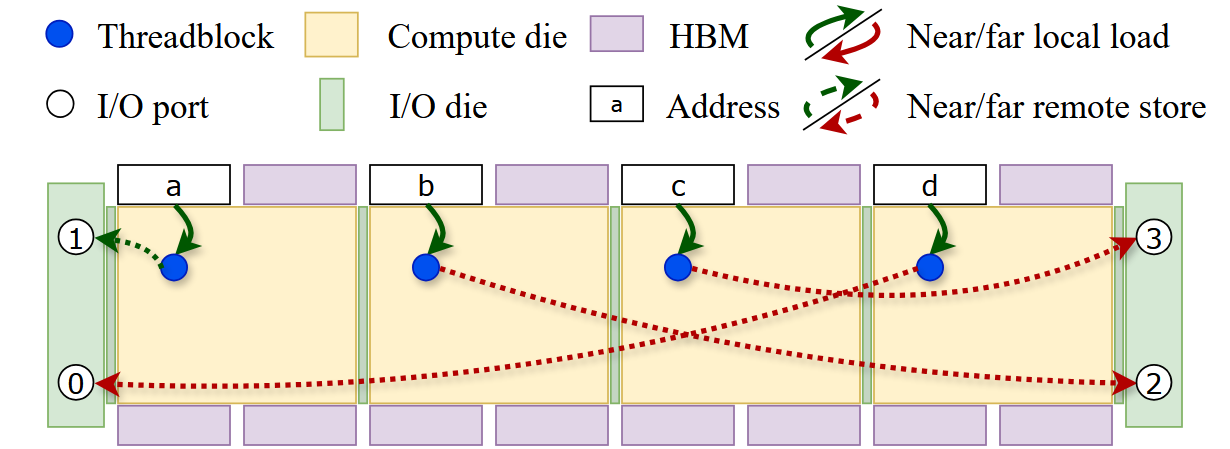}
        \vspace{-.2em}
        \caption{NUMA-aware placement with baseline routing}
        \label{fig:all_traffics_numa}
        \vspace{.1em}
    \end{subfigure}

    \begin{subfigure}{\linewidth}
        \centering
        \includegraphics[width=0.85\linewidth]{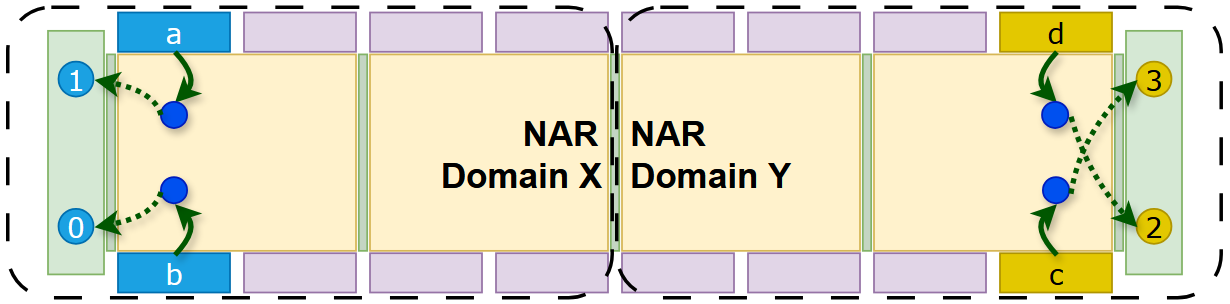}
        \vspace{-.2em}
        \caption{NUNA-aware placement with \nar{}}
        \label{fig:all_traffics_narnap}
    \end{subfigure}

    \vspace{-.2em}
    \caption{Example of (a)~baseline and (b)~granularity-2 \nar{} address to I/O port mappings.
    (c)~An example of baseline and \nar{} routing for local loads (solid arrows) and remote stores (dashed arrows) between threadblocks and data for NUMA-aware placement,
    and (d) \nap{} with \nar{}.
    }
    \label{fig:all_traffics}
    \vspace{-1.2em}
\end{figure}

\subsubsection{Improved I/O Routing Policy}\label{sec:nuna_aware_routing_nar}
\nar{} routes latency-sensitive traffic to only a subset of I/O ports physically closer to the CU, referred to as a~\emph{\nar{} domain}.
To balance latency and bandwidth, we introduce~\emph{\nar{} granularity}, defined as the number of available~\nar{} domains.
Higher \nar{} granularity denotes fine-grained \nar{} group, each with smaller number of I/O ports. This scheme helps each CU route inter-GPU flits to a very specific subset of I/O ports physically close to it, reducing the latency.
However, the number of I/O ports each CU can utilize is limited, hurting the network bandwidth for larger collectives.
\nar{} leverages a statically partitioned physical address space, as illustrated in~\autoref{fig:phys_addrs_and_nar}. The physical address space is divided into distinct segments, each representing to a specific \nar{} granularity (e.g., granularity-2, granularity-4, etc.).
Each segment utilizes all I/O ports, partitioned uniformly into subsets for each domain.

To implement \nar{}, we propose modifying NoC flit generation logic.
The upper address bits determine the segment.
Each segment may utilize different \nar{} granularity or even baseline routing, providing routing flexibility.
The number of I/O ports in each \nar{} domain is the total number of ports divided by granularity.
For example, in~\autoref{fig:phys_addrs_and_nar}, the socket has four I/O ports; each \nar{} domain has two I/O ports when the granularity is two.
Each~\nar{} domain statically maps to a subset of ports, via offset. The yellow domain in~\autoref{fig:phys_addrs_and_nar} has an offset of two to target I/O ports 2--3.
Within a \nar{} domain, we load balance across I/O ports.
This requires no changes to the NoC.
Once the I/O port is chosen by \nar{}, the on-chip transfer continues as normal.

Note this physical address partitioning scheme also facilitates page-to-HBM-stack placement implementation.
The physical addresses for a~\nar{} domain will not only be restricted to a subset of I/O ports, but also to a specific HBM stacks closer to the subset of I/O ports.
\autoref{sec:nuna_aware_placement} articulates the placement policy in detail.

New extensions to backend toolchains will allow specifying the \nar{} policy and allocate memory to the statically partitioned address ranges.
The information is then stored in the page table and retrieved when the GPU accesses memory. 
Pages allocated to specified \nar{} granularities and domains simply need to have the correct virtual-to-physical translation, and hardware will handle off-chip routing accordingly.
Within a~\nar{} domain, remote traffic is still load-balanced amongst the subset of I/O ports.
This segmentation of the address space will have a negligible impact on the overall GPU memory capacity because we target small collectives and only a few buffers; the size of segments is much smaller (e.g., 100s of MB) than the entire physical address space (e.g., 100s of GB). Furthermore, the size of the segments can be determined at boot to adjust to communication demand.

To maximize benefit, \nar{} must avoid hotspotting on the I/O ports.
Generally, for latency-bound collectives, load balancing to improve bandwidth is not a concern but for large collectives.
In fact, as evaluated in~\autoref{sec:results}, the speedups from \nar{} diminish for larger collectives.
Thus, we rely on the compiler to select whether to use \nar{} for a collective or leverage the baseline routing scheme for better load balancing.
It may exhaustively try all available \nar{} granularities (the algorithm runs quickly) or use heuristic thresholds\footnote{For individual GPU architectures, the optimal collective size threshold for using \nar{} can be statically determined.}.

\insertFigureWide{Allocation_Diagram}{fig:alloc_alg_visual}{0.9}{-1.3em}{-1.6em}{
Visual example of logical allocation and physical placement steps for a two-GPU, two-channel \allgather{} implemented as a local load, remote store using a single buffer. 
The threadblocks~\texttt{w-z} in GPUs~\texttt{0} and~\texttt{1} are locally loading (solid arrow) and remotely storing (dashed arrow) from chunks~\texttt{a-d}.
In \texttt{Logical Allocation}, threadblocks and local/remote chunks are logically grouped~(colored) and allocated to \nar{} domains~\texttt{0-3} in round-robin fashion.
In \texttt{Physical Placement}, the threadblocks are mapped to CUs and pages backing chunks are mapped to HBM stacks.
For example, threadblock~\texttt{w} on GPU~\texttt{0} performs a near local load and a remote store using the shortest path(s) through \nar{} domain ~\texttt{0} }

\subsection{\ours{}-Aware Placement}\label{sec:nuna_aware_placement}

At a high level, \nap{} not only places threadblocks and memory closer to themselves; we put them to the CUs and HBM stacks closer to the I/O ports (\ie{} outwards to the socket) to minimize the inter-socket latency.
In~\autoref{fig:all_traffics}, we illustrate our two proposed \ours{}-aware optimizations, \nar{} and \nap{}.
\autoref{fig:all_traffics_numa} shows NUMA-optimized threadblock scheduling and memory placement utilizing the baseline off-chip routing (\autoref{fig:nar_diagram_baseline_hashing}).
In the baseline, addresses \texttt{a}--\texttt{d} are mapped one-to-one to any of the ports 1--4, achieving perfect load balancing across I/O ports. 
While an example NUMA-optimized collective assures near local loads for all chunks, it may route a flit to far I/O ports due to na\"ive hashing (e.g., \texttt{d}) or the placement of some buffers/chunks (e.g., \texttt{c}). %
These unnecessarily long paths (red dashed arrows) (\rom{1})~increase individual latency of communication and (\rom{2})~increase congestion in the middle of the NoC.
\autoref{fig:nar_diagram_nar_hashing} shows how \nar{} modifies the I/O port hashing scheme so that each illustrated \nar{} domain utilizes only a subset of ports closer to the CU.
In~\autoref{fig:all_traffics_narnap}, we combine \nar{} with \nap{} to schedule threadblocks and place memory to optimal CUs and HBMs, respectively.
Addresses \texttt{a} and \texttt{b} are placed in~\nar{} domain \texttt{X} and similarly for \texttt{c} and \texttt{d} in~\nar{} domain \texttt{Y}.
A \ours{}-aware collective uses \nar{} for (nearest) I/O port and \nap{} for threadblocks and buffers to achieve near local and remote operations.

The \nap{} placement algorithm is provided in~\autoref{alg:allocation} and a visual example of the allocation algorithm for a two-GPU \allgather{} is illustrated in~\autoref{fig:alloc_alg_visual}.
The compiler/user specifies the collective communication schedule ($W$) and \nar{} granularity ($nar_{max}$) as inputs. The nearest HBM stack for each CU (derived from $topology$) and maximum resources available ($R_{max}$) are assumed to be known for an architecture. %
The \ttvar{logical_allocation} phase decomposes the target collective schedule and groups threadblocks and their associated chunks into logical groups.
After the logical allocation, the \ttvar{physical_placement} phase iterates over each logical domain and greedily places the threadblock(s) and chunk page(s) in the order physically closer to the I/O ports, keeping track of resource usage/contention.
This process maintains compute-to-memory locality by assigning related threadblocks and buffers to the same logical group. %

For simplicity, we assume the communication schedule is explicitly provided via a collective plan (e.g., MSCCLang~\cite{cowan2023mscclang}), providing exact threadblock access patterns. Alternatively, if explicit plans are unavailable, these access patterns can be detected automatically from kernel code via static compiler analysis~\cite{kim2017coda,khairy2020locality,cabezas2015automatic,li2017locality,chen2017improving,kim2023locality,zhu2024spgpu}.
The plan statically defines \emph{chunks} as fixed-size regions with a GPU buffer uniquely identified by the tuple~\emph{(chunk\_id, buffer\_type, gpu\_id)}.
\nap{} colocates threadblocks that communicate with the same chunks via ~\ttvar{logical_allocation} and places these threadblocks/chunks in appropriate \nar{} segments via~\ttvar{physical_placement}.
We use 4\,kB pages, allowing \nap{} to control placement for even the smallest of latency-sensitive collectives.

\subsubsection{Logical Allocation}\label{sec:nuna_aware_placement_logical_allocation}
\ttvar{logical_allocation} assigns threadblocks and chunks into logical groups, resolving access dependencies before physical placement to prevent order-dependent fragmentation. Because hardware resources ($R_{max}$) are limited within each \nar{} domain, performing \ttvar{physical_placement} directly could split related groups across the die if they exceed resource budgets during sequential processing. By first partitioning the workload into logical groups,~\nap{} can load-balance them across the GPU, ensuring related tasks remain colocated. %

Collective decomposition identifies access patterns.
Next, \ttvar{logical_allocation} processes threadblocks sequentially to group them and their associated chunks into~\emph{logical groups}---threadblocks and chunks intended to be co-located on the same~\nar{} domain and nearby CU/HBM.
If a threadblock's chunks are unassigned, it forms a new logical group; otherwise, the threadblock and chunks join the existing group with which it shares the most chunks to maximize overlap.
This greedy policy prioritizes colocation and may produce groups of varying sizes depending on access patterns. \ttvar{logical_allocation} returns a list of logical groups, $L$, sorted in decreasing order of size, naturally load balancing subsequent~\ttvar{physical_placement}.

In practice, well-structured collective plans tend to minimize overlap. Many threadblocks access mutually exclusive chunk tuples (e.g., disjoint regions of output/input/scratch per GPU) through independent communication streams. %
Thus, logical allocation often produces domains that are perfectly or nearly disjoint in chunk access (\eg{}~\autoref{fig:alloc_alg_visual}), and cross-domain accesses are rare.
We handle conflicts deterministically in~\ttvar{physical_placement}.

\begin{algorithm}[]
\scriptsize
\caption{\ours{}-Aware Placement} \label{alg:allocation}
\begin{algorithmic}[1]
\State \textbf{Given:} $topology$ \Comment{CU, HBM, and I/O locations in topology}
\State \textbf{Given:} $R_{max}$ \Comment{Resource constraints}

\State \textbf{Input:} $nar_{max}$ \Comment{\# \nar{} domains}
\State \textbf{Input:} $W = \{ (gpu,tb,C_{tb}),\ldots \}$ \Comment{Workload description}
\State \textbf{Output:} $M_{compute} : tb \mapsto cu$ \Comment{Threadblock to CU assignment}
\State \textbf{Output:} $M_{memory} : C_{tb} \mapsto stack$ \Comment{Chunk to HBM assignment}

\State $\forall cu, \quad R[cu] \gets R_{max}$
\State $\forall stack, \quad R[stack] \gets R_{max}$
\State $nar \gets 0$

\State $L \gets $ \func{logical\_allocation}($W$) %
\For{$L_i \in L$}
    \State $nar \gets (nar + 1) \mod nar_{max}$  \Comment{Round-robin \nar{} domains}
    \For{$(tb_{j},C_{j}) \in L_i$}
        \State $cs \gets |C_{j}|$ \Comment{Chunk size}

        \\ \Comment{\ours{}- and state-aware}
        \State $cu, stack \gets $ \func{physical\_placement}($nar$,$cs$,$R$, $topology$)
        \State $R[cu] \gets R[cu] - cs$  \Comment{Update CU resource}
        \State $R[stack] \gets R[stack] - cs$  \Comment{Update memory resource}
        \State $M_{compute}[tb_{j}] \gets cu$
        \State $M_{memory}[C_{j}] \gets stack$

    \EndFor
\EndFor
\end{algorithmic}
\end{algorithm}

\subsubsection{Physical Placement}\label{sec:nuna_aware_placement_physical_placement}
Given the \nar{} domains, the physical placement process calculates the threadblock-to-CU ($M_{compute}$) and chunk-to-HBM ($M_{memory}$) mappings.
\autoref{fig:alloc_alg_visual} shows an example of mapping the logical groups onto two \nar{} domains.

\paratitle{CU and stack selection under resource constraints}
In~\autoref{alg:allocation}, each logical group is assigned to a NAR domain, $nar$ (line 10).
Round-robin allocating logical groups to \nar{} domains (line 12) implicitly balances I/O port usage/load.
The function \ttvar{physical_placement} returns the chosen CU and HBM stack, $cu$ and $stack$, for each threadblock and buffer in the logical group (line~16).
For each threadblock, it selects the first CU whose incremental load fits the CU resource budget and an HBM stack, preferring the CU's nearest stack.
The scheme can be extended to select a set of CUs if CU resource contention is expected to be high.
In~\autoref{alg:allocation}, $R_{max}$ specifies the limiting resource(s) for CUs and memory. For simplicity of representation, we show the resource tracking variable, $R$, as keying off the resource type (CU or HBM stack IDs) and accepting chunk size ($cs$) in bytes inputs that are translated to the correct resource (outstanding requests or bandwidth).
For CUs, $R_{max}$ represents the maximum outstanding request capacity (e.g., 700 requests per CU).
For HBM stacks, $R_{max}$ is calculated based on the bandwidth-delay product, estimating the volume of data that can be in flight before saturation.
After each decision, the resource state is updated (lines~17--18), and $M_{compute}$ and $M_{memory}$ are recorded (lines~19--20).

\paratitle{Page-level placement and conflicts}
The first chunk that triggers the page placement fixes the page's stack, and (potential) subsequent chunks in the same page may incur a non-ideal placement.
If multiple chunks would prefer different stacks for the same page, we resolve the conflict by first-come, first-served in the \ttvar{logical_allocation} processing order.
For sub-page chunks, we pack multiple chunks into the first-chosen page, consistent with a page-based virtual-to-physical translation.

\subsubsection{Implementing \nap{}}
\label{sec:nap_implementation}

\autoref{alg:allocation} outputs two logical maps: threadblock-to-CUs ($M_{compute}$) and chunk-to-stack ($M_{memory}$).
Realizing these allocations in hardware requires controlling two hardware mechanisms: the CU(s) that run a threadblock and the stack that holds a page.
Existing hardware APIs can implement the threadblock allocation, with changes similar to those in existing academic and industry works allowing control of page-to-stack placement.

\paratitle{Threadblock Dispatch}
The dispatcher launches threadblocks to a subset of preferred CU(s) as specified through existing APIs. 
Existing hardware exposes scheduling at the CU granularity~\cite{amd2025hip,nvidia2025CUDAToolkit}. Static control of threadblock placement at this granularity is also a standard assumption in prior GPU locality works~\cite{vijaykumar2018locality,zhu2024spgpu}.
Assuming hardware support, the only requirement is passing GPU-specific topology and CU mappings (i.e., $topology$ in ~\autoref{alg:allocation}) to the compiler.
Contention between existing threadblocks and \nap{}/collective threadblocks can be resolved by ignoring the CU affinity and defaulting to the baseline dispatch policy.

\paratitle{Memory Placement}
The driver's memory manager realizes \\$M_{memory}$ by mapping the virtual address(es) of a chunk's page(s) onto the physical addresses backed by the desired stacks. We assume that \nar{} domains are statically partitioned into per-stack regions as illustrated in~\autoref{fig:phys_addrs_and_nar}.
These partition boundaries can be configured at boot via firmware modifications to the data fabric routing and memory manager.
The communication backend toolchain/compiler decides memory placement using the stack-to-I/O spatial relationship (i.e., $topology$ in~\autoref{alg:allocation}). Using that spatial mapping, the driver implements the desired placement policy without making any changes to address translation or the memory controller.
Partitioning the physical address space into hardware-partitioned memory domains has existing implementations in GPUs (AMD NPS~\cite{amdDeepDive,tee2025mall} and NVIDIA MIG~\cite{nvidiamig}) and
CPUs (Linux OS memory drivers~\cite{verghese1996osnuma,lameter2013numa}).
Prior locality works control placement at the data structure~\cite{vijaykumar2018locality}, region~\cite{zhang2022sdam}, or page~\cite{hsieh2016tom} granularity.

\section{Methodology}\label{sec:methodology}

\insertFigure{NoC_Fabric_Diagram}{fig:noc_network}{0.7}{0em}{-0em}{
Evaluated on-chip network.
24$\times$6 mesh connects the CUs in compute dies (yellow).
Mesh routers in vertical edge connect to 12 I/O ports (green).
Routers in the horizontal edges connect to HBM stacks (purple) with full connectivity to all 16 memory channels within each stack.
}

\subsection{Simulation Infrastructure}\label{sec:simulation_infrastructure}

We evaluate collective communication performance using the ASTRA-sim simulator, which is real-system correlated and widely adopted~\cite{llmservingsim2, multiverse, simai}.
Specifically, we use ASTRA-sim~3.0~\cite{won2026astrasim3} that models threadblock-level operations at the CU-level.
ASTRA-sim~3.0's GPU model simulates NoC and scale-up network transactions at the 256\,B cache-line-sized flits granularity.
Each compute and reduction operation consumes CU resources (\ie{} blocking other threadblocks) using a delay model based on hardware floating point operations per second~(FLOPS) specifications.
Also, we add a constant 120\,ns latency for each memory access.
This is because we evaluate collective communications without data reuse, resulting in cache misses.
Finally, we evaluate latency-optimized collective algorithm (i.e., direct/\allpairs{}) with the best-performing number of threadblocks, using threadblock-level MSCCLang collective representation~\cite{cowan2023mscclang}.

\subsection{End-to-End LLM Evaluation Methodology }
\label{sec:method_llm}

For end-to-end evaluations, we target 12 dense and sparse (\ie{}~MoE) LLM architectures:
Llama (7B/70B/405B) \cite{meta2025llama,grattafiori2024llama3}, GPT-OSS (20B/120B) \cite{openai2025gptoss}, Mixtral (8$\times$7B/8$\times$22B) \cite{jiang2024mixtral}, Qwen3 (30B-A3B/235B-A22B)~\cite{yang2025qwen3}, and DeepSeek (V3.2/V4-Flash/V4-Pro) \cite{deepseekv3,deepseekv4}. %
We used an in-house tool to generate forward pass inference trace (tensor shapes, kernel signatures, and communication events) for each input model, hardware, and parallelization configurations.
Using system parameters from~\autoref{tab:simul_params}, the tool calculates compute time through a roofline model.
Similarly, using NoC and scale-up network parameters, it estimates the collective communication time. Note that we model compute and communication serially since intra-batch overlap is ongoing research~\cite{gond2025tokenweave,li2024tpi,patel2024splitwise,pati2024t3}.
We choose the batch size, sequence length, and parallelization strategy that minimizes time-to-first-token (TTFT) for prefill and time-per-output-token (TPOT) for decode.

\subsection{Evaluated System and Configurations}\label{sec:method_system}

\begin{table}[t]
\centering
\caption{Simulated system parameters.}
\vspace{-.4em}
\label{tab:simul_params}
\begin{tabular}{cl}
\toprule
\textbf{Config.} & \textbf{Value} \\ \midrule
Dies & \begin{tabular}[c]{@{}l@{}}4$\times$ compute, 2$\times$ I/O per GPU\\ 576 CUs, 700 outstanding requests per CU\\ 120~ns (L1) latency from CU to NoC\\
100\,PFLOPS fp4, 25\,PFLOPS fp/bf16
\end{tabular} \\ \hline
NoC & \begin{tabular}[c]{@{}l@{}}24$\times$6 mesh @ 1.2GHz\\ 10~TB/s bisection bandwidth 10 cycle hop latency\\ DOR (X-Y) routing\end{tabular} \\ \hline
Memory & \begin{tabular}[c]{@{}l@{}}40~TB/s cumulative bandwidth, 150~ns access latency\\ 16 HBM stacks\end{tabular} \\ \hline
Switch & \begin{tabular}[c]{@{}l@{}}Single level Clos\\ 144~I/O ports (72 per die)\\ 7.2~TB/s cumulative bandwidth \\500~ns die-to-die latency\end{tabular} \\
\bottomrule
\end{tabular}
\vspace{-0.8em}
\end{table}

We evaluate a scale-up pod of multi-die GPUs\footnote{In evaluation, ``number of GPUs'' refers to the number of sockets. Thus, the number of individual dies evaluated is four times that value.} discussed in~\autoref{sec:motivation}.
\autoref{tab:simul_params} shows system parameters, based on previous work~\cite{jin2024uncovering,luo2024benchmarking,arunkumar2017mcm,dalmia2024cpelide} and industry trends~\cite{amd2025advancingai,nvidia2025gtc,naddodNVIDIAGB200}.
\autoref{fig:noc_network} depicts the target NoC, compatible with the different design philosophies and profiling of current architecture.
We evaluate a global 24$\times$6 mesh over four dies with X-then-Y routing~\cite{dally2004principles}. The hop latencies are scaled to match measured values in real hardware~\cite{jin2024uncovering,luo2024benchmarking,arunkumar2017mcm,dalmia2024cpelide}.
Note our profiling of current hardware shows fine-grained latency differences consistent with a multi-hop network.

The scale-up network is a single-level Clos as described in~\autoref{fig:nuna_evaluation_system}.
The baseline system hashes and distributes inter-GPU traffic across all I/O ports for load balancing.
\nar{} is implemented as described in~\autoref{sec:nuna_aware_routing}, utilizing a subset of I/O ports based on the physical address (e.g.,~\autoref{fig:phys_addrs_and_nar}).
We model the threadblock scheduler, memory management unit, and virtual address space after modern shared memory GPU systems.
The default threadblock allocation policy schedules threadblocks across dies in a round-robin order, prioritizing free CUs for load balancing.
The baseline round-robin allocates pages to HBM stacks.
To model~\numa{} optimizations for local data access~\cite{vijaykumar2018locality,khairy2020locality}, we assume baseline memory placement but run~\autoref{alg:allocation} for threadblock scheduling near associated memory.

We evaluate multiple configurations:
\begin{itemize}[leftmargin=*]
    \item \unaware{}: \ours{}-unaware baseline. Threadblocks and memory pages are mapped using round-robin placement and off-chip traffic hashes across all I/O ports.
    \item \unaware{}+\nar{}: \nar{} is applied to round-robin threadblock/memory placement.
    \item NUMA: Proxy to NUMA-aware optimizations with baseline memory allocation and nearby CU scheduling~\cite{vijaykumar2018locality,khairy2020locality}.
    \item NUMA+\nar{}: \nar{} is applied to the NUMA threadblock and memory placement.
    \item \nap{}: \ours{}-aware threadblock and memory placement, but traffic is hashed across all I/O ports.
    \item \napnar{}: Fully \ours{}-aware configuration, integrating both \nar{} and \nap{} techniques.
    \item \nap{}*: \ours{}-aware threadblock placement while employing the baseline round-robin memory placement policy.
\end{itemize}

\begin{figure}[t]
    \centering
    \begin{subfigure}{\linewidth}
        \centering
        \includegraphics[width=0.95\linewidth]{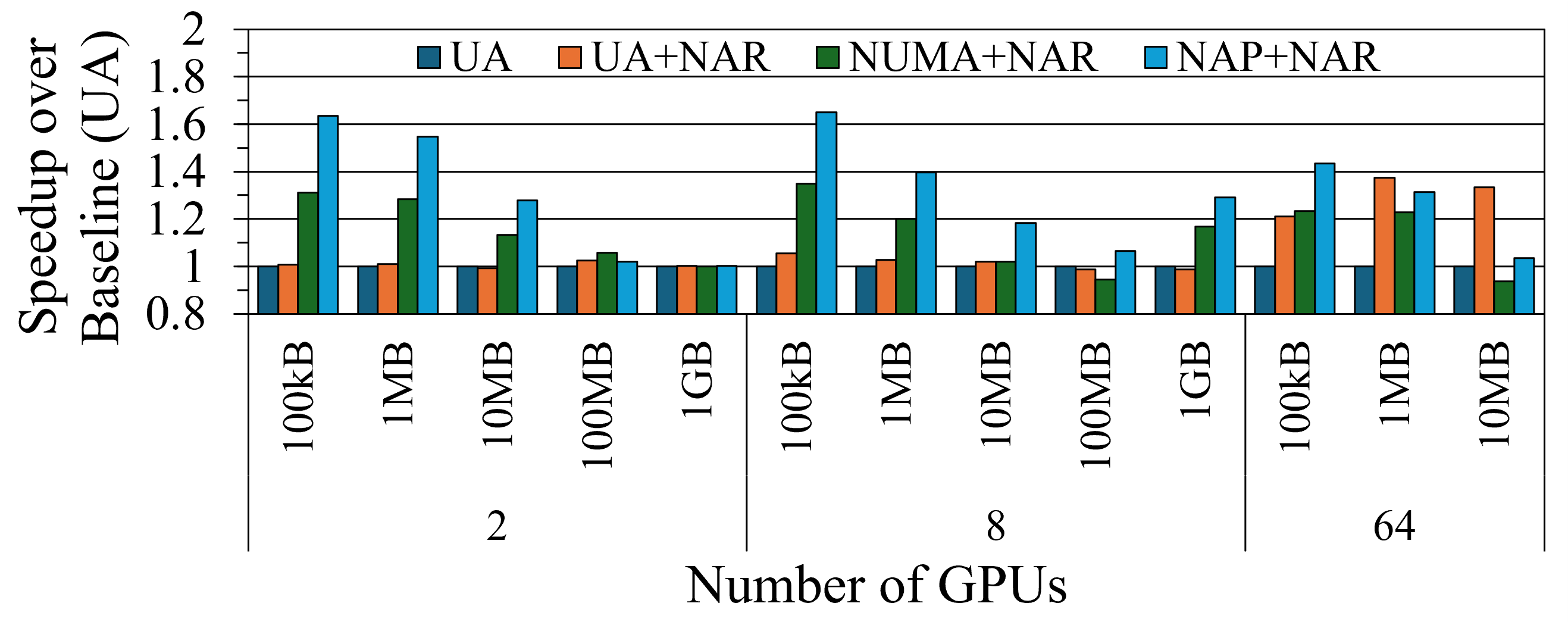}
        \vspace{-0.4em} %
        \caption{\allgather{}}
        \label{fig:topline_ag}
    \end{subfigure}
    \begin{subfigure}{\linewidth}
        \centering
        \includegraphics[width=0.95\linewidth]{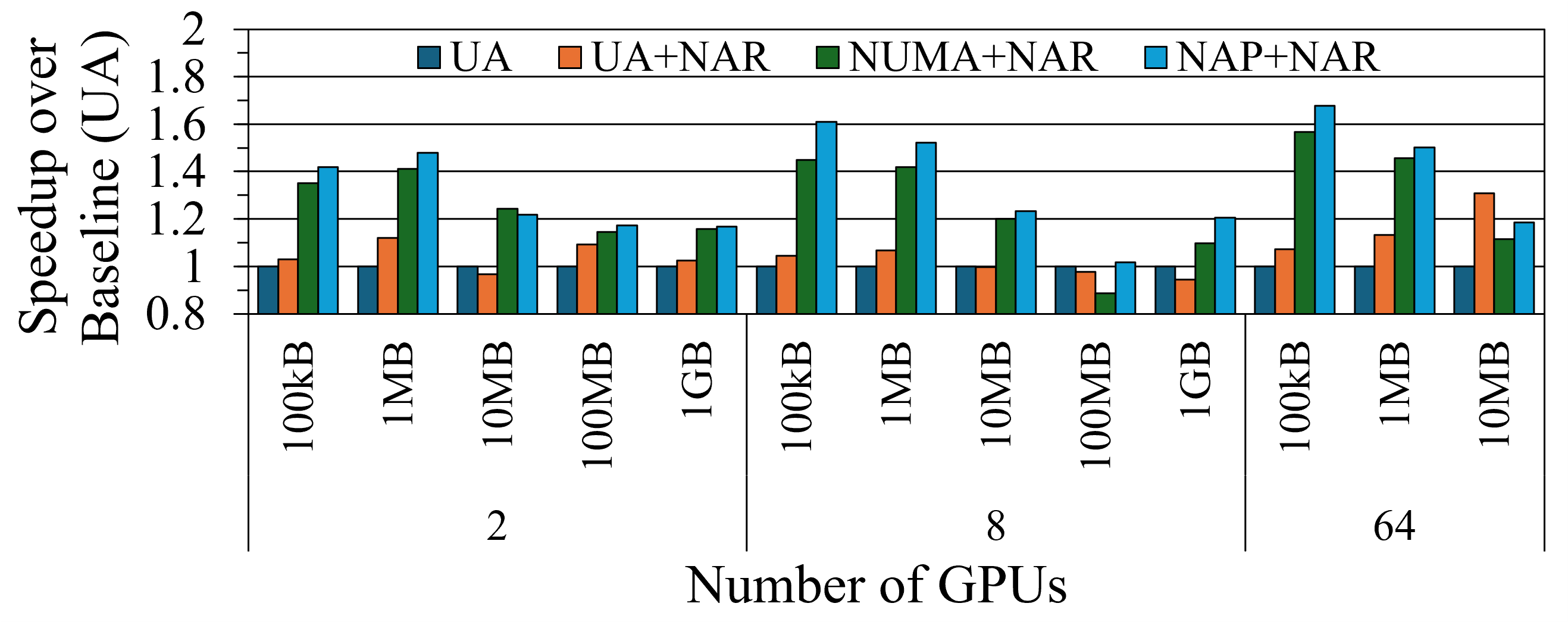}
        \vspace{-0.4em} %
        \caption{\allreduce{}}
        \label{fig:topline_ar}
    \end{subfigure}

    \begin{subfigure}{\linewidth}
        \centering
        \includegraphics[width=0.95\linewidth]{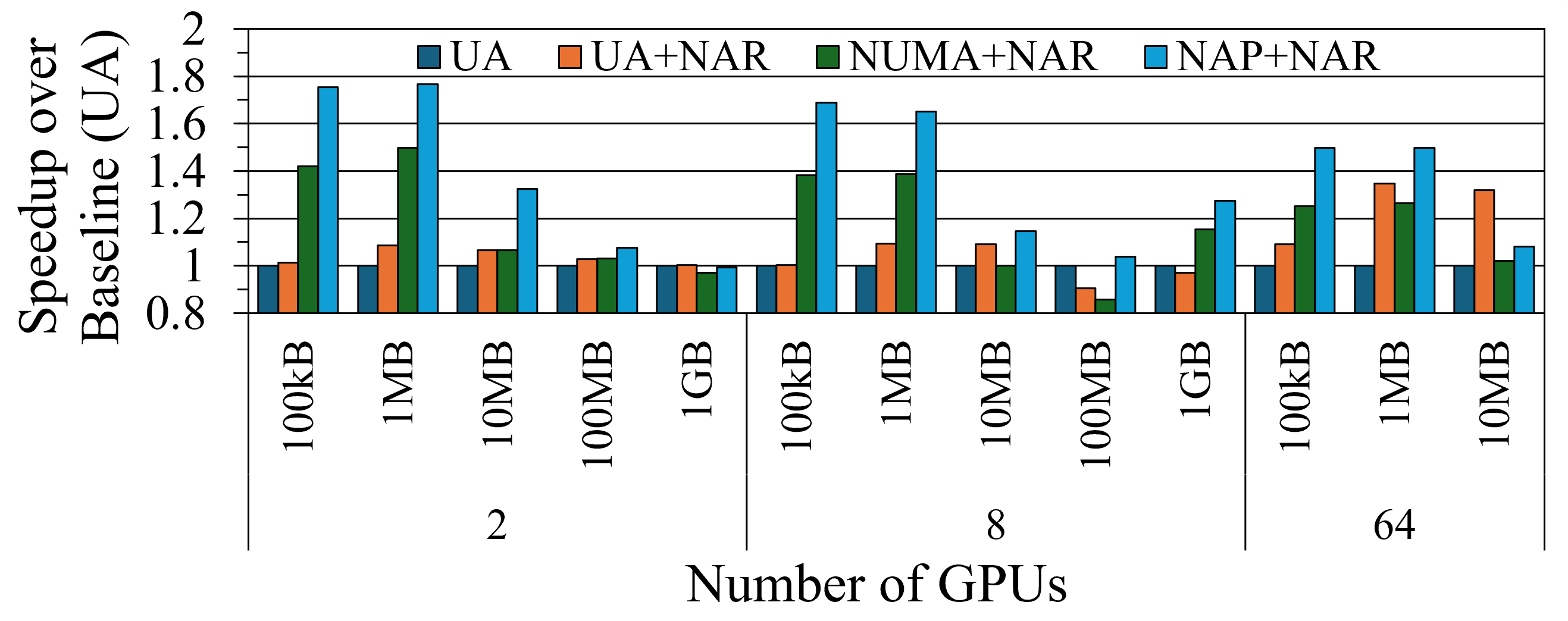}
        \vspace{-0.4em} %
        \caption{\alltoall{}}
        \label{fig:topline_a2a}
    \end{subfigure}
    \vspace{-1.5em}
    \caption{Execution speedups across pod and collective sizes for \allgather{}, \allreduce{}, and \alltoall{}.}
    \label{fig:topline}
    \vspace{-0em}
\end{figure}

\section{Results}\label{sec:results}

\subsection{Microbenchmarks}
\label{sec:results_micro}

We measured the speedup of \ours{}-aware techniques over the baseline~(\unaware{}),
to determine the utility of spatial optimization. %
We evaluated~\allgather{},~\allreduce{}, and~\alltoall{} collectives ranging from 100\,kB to 1\,GB, across scale-up pods of size 2--64~GPUs.
\autoref{fig:topline} plots the relative speedups. %

For small collectives, the speedups of \ours{}-aware techniques remain consistent across both collectives and pod size configurations. There is a consistent trend where \nar{} alone improves performance over the \unaware{} baseline.
NUMA+\nar{} increases speedup by reducing local latencies to I/O ports.
Finally, \napnar{} is superior by reducing both local and remote latencies. As expected, the smaller, latency-bound collectives see the highest speedups.
Notably, the lower speedup of \napnar{} for 64-GPU, 10\,MB collectives over the \unaware{}+\nar{} configuration is due to the oversubscription of each \nar{} domain.
Threadblocks accessing the same chunk are collocated into a~\nar{} domain so large clusters with many threadblocks oversubscribe the resources in each~\nar{} domain.

\begin{figure}[t]
    \centering
    \begin{subfigure}[b]{0.49\linewidth}
        \centering
        \includegraphics[width=\linewidth]{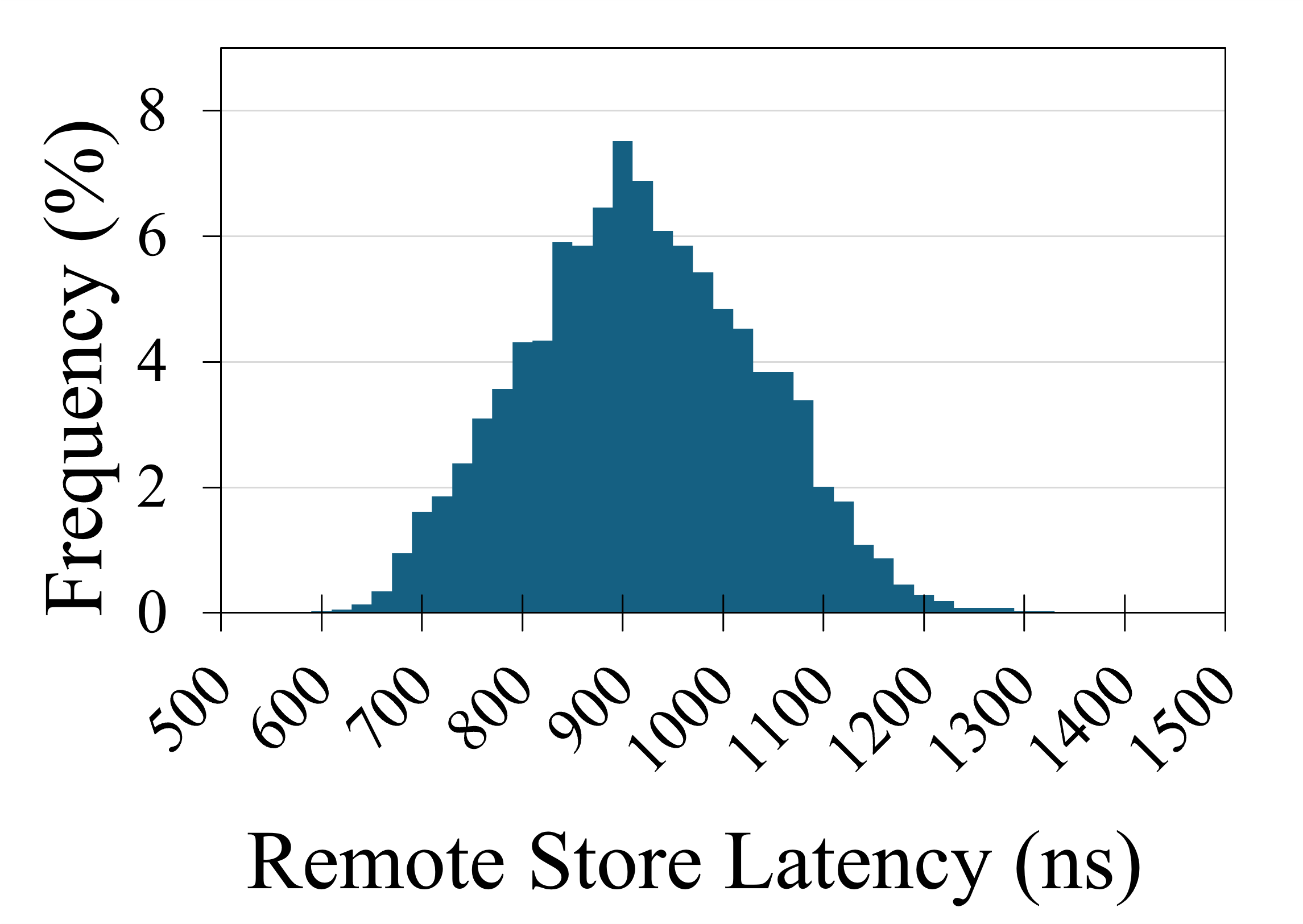}
        \vspace{-1.4em} %
        \caption{\unaware{}}
        \label{fig:latency_histo_sua}
    \end{subfigure}
    \hfill
    \begin{subfigure}[b]{0.49\linewidth}
        \centering
        \includegraphics[width=\linewidth]{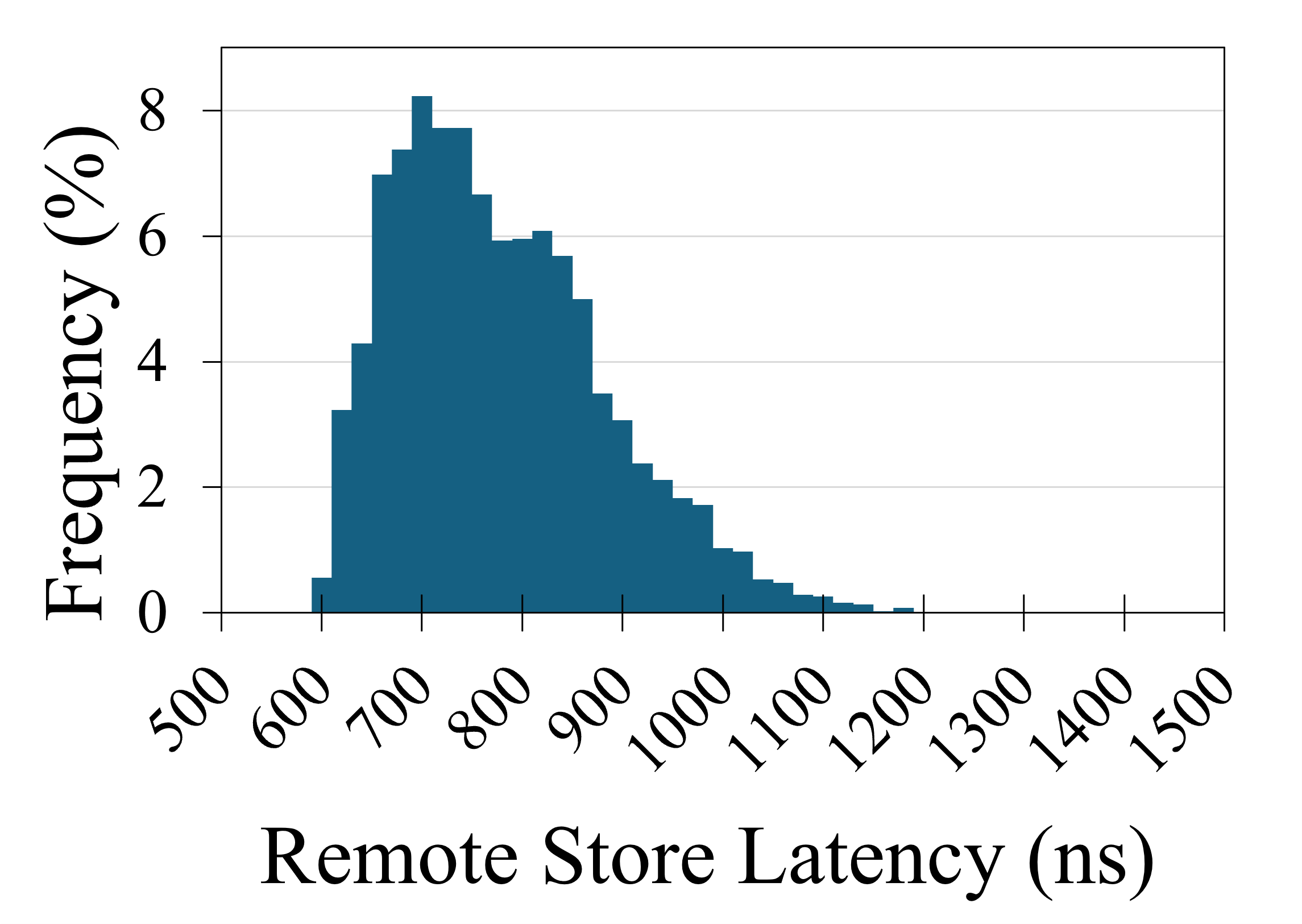}
        \vspace{-1.4em} %
        \caption{\napnar{}}
        \label{fig:latency_histo_scall}
    \end{subfigure}
    \vspace{-1.4em} %
    \caption{Latency distributions of remote stores in an eight-GPU, 1\,MB \allgather{} for baseline (\unaware{}) and (\napnar{}).
    }
    \label{fig:latency_histo}
    \vspace{-.6em}
\end{figure}

\subsection{Latency Distributions}

To further understand how \napnar{} mitigates intra-socket non-uniformity for inter-GPU communications, we measured the latency distributions of one-way, off-chip stores. Specifically, we evaluated an eight-GPU cluster with 1\,MB \allgather{} collectives for both \unaware{} and \napnar{} configurations. The latency distributions are illustrated in~\autoref{fig:latency_histo}.

The \unaware{} baseline exhibits higher average and maximum latencies compared to the \napnar{} configuration. The \unaware{} baseline shows 920\,ns and 1,320\,ns for measured average and maximum latencies, respectively.
\napnar{} records 780\,ns and 1,180\,ns, respectively. The bell-shaped curve of the unaware baseline histogram in~\autoref{fig:latency_histo_sua} represents the Gaussian distribution of individual path latencies in a mesh, demonstrating symmetric non-uniformity. In contrast, the \ours{}-aware configuration in~\autoref{fig:latency_histo_scall} shows a positively skewed normal curve by reducing both local and remote access latencies. The long tail indicates congestion at the bottlenecking I/O ports, but the maximum latency is still 10\% lower than that of the unaware configuration.

\insertFigure{isolations}{fig:isolations}{1}{-1.0em}{-0em}{
Isolation of techniques for small (100\,kB) and medium (10\,MB) \allgather{} for eight GPUs.
}

\subsection{Isolation of Techniques}

To determine the individual impact of the \ours{}-aware techniques, we evaluated multiple configurations where \nar{} and \nap{} were run in isolation. A small (100\,kB) and a medium (10\,MB) \allgather{} were run for an eight-GPU configuration.
Relative speedups over the \unaware{} baseline for each configuration are given in~\autoref{fig:isolations}.

Notably, for all threadblock and memory placement schemes (\unaware{}, NUMA, \nap{}*, and \nap{}), \nar{} improves performance.
Among the different placement policies, \nar{} has the greatest impact on \nap{}.
For 10\,MB collectives, placement policies (NUMA, \nap{}*, and \nap{}) alone without \nar{} introduce slowdowns due to NoC congestion caused by a large volume of hashed traffic without \nar{}.

Furthermore, the isolation study highlights the importance of \nar{} in conjunction with \nap{}.
The baseline placement evenly distributes threadblocks across a socket, causing traffic to experience average-case latencies.
However, isolated \nap{} is not beneficial.
Although threadblocks are placed outwards, as traffics are evenly hashed, many flits utilize I/O ports on the opposite edge and experience worst-case internal latency.
This result demonstrates placement alone is not enough for optimizing inter-GPU communications; to mitigate the \ours{} effect, both placement and routing should be jointly considered.

\subsection{Sensitivity to Collective Size}

\insertFigure{size_sweep}{fig:size_sweep}{0.95}{-.6em}{-0.4em}{
Speedup of \napnar{} over the baseline (\unaware{}), and \napnar{} scale-up network bandwidth utilization, for two-GPU \allgather{} across distinct output buffer sizes.
}

Expanding on the claim that \ours{} optimization is most effective for latency-bound collectives, we measured the effect of collective size on the \napnar{} configuration.
\autoref{fig:size_sweep} captures the result.
\ours{}-aware \napnar{} speedup benefits are greater when the collectives are small, as the inter-GPU bandwidth is not saturated and transmission times are closer to link latencies, making them sensitive to intra-socket non-uniformity. However, as the collective size increases, the speedup
diminishes. When the inter-GPU network becomes saturated, intra-socket non-uniformity is hidden by the dominant long queuing delays at the scale-up NICs and switches.

\subsection{Sensitivity to \nar{} Granularity}

We also measured the effect of \nar{} granularity on the \nar{} routing technique.
We evaluated 1\,MB \allgather{} collectives with two configurations: one with high parallelization (1,008 threadblocks per GPU) and the other with low parallelization (7 threadblocks per GPU).
The speedups of 2--12~\nar{} domains over baseline routing are plotted in~\autoref{fig:nar_sensitivity}.
As expected, when there are a sufficient number of threadblocks to enable effective load balancing, the configuration with the finest granularity (12 \nar{} domains) achieves the lowest latency and performs the best. However, when there are too few threadblocks to distribute across many \nar{} domains, traffic becomes unequally routed, resulting in slowdowns.
This trade-off between low latency and load balancing is implicitly considered in the algorithmic \nap{} allocation, which determines the best number of domains based on the workload.

\insertFigure{nar_sensitivity}{fig:nar_sensitivity}{0.95}{-.2em}{-0em}{
The speedup of \nar{} (\unaware{}+\nar{}) over the baseline (\unaware{}), over different \nar{} granularities, for an eight-GPU 1\,MB \allgather{}. The number of threadblocks was set to a high value to fill all CUs (many) or just a couple (few).
}

\insertFigure{gemm_sweep_rel}{fig:gemm_sweep}{1}{-.2em}{0em}{
Impact of concurrent GEMM kernel on unaware baseline (\unaware{}) and \ours{}-optimized configuration (\napnar{}).
Execution time relative to baseline (UA) of an eight-GPU, 1\,MB \allgather{} was measured for a sweep of matrix sizes.
}

\subsection{Interactions with Normal Execution} \label{results_gemm}

While these optimization techniques work well in isolation, they might be susceptible to forming hotspots or under-utilization when used with other kernels.
To measure this effect, we evaluated the performance of \ours{} techniques with an independent general matrix-multiply~(GEMM) operation with 100 tiles.
Its threadblocks were naively allocated across the GPU.
They load the required data from input matrices, perform a computation as a static delay based on data size, and store the result to the output matrix.
An eight-GPU, 1\,MB \allgather{} was concurrently run. 
We measured the collective speed over a sweep of GEMM sizes for unaware (\unaware{}) and aware (\napnar{}) to see when or if the collective performance degrades.
\autoref{fig:gemm_sweep} summarizes the result.

Significant on-chip GEMM traffic congests the NoC and causes the collective communication execution speed to degrade sharply.
The \napnar{} configuration saturates at a higher GEMM size, implying that the \ours{}-aware configuration is more robust to concurrent execution.
\ours{} techniques localize the access to the I/O ports and minimize the traffic in the middle of the NoC.
Therefore, the solution is more robust to the NoC congestion. The collective performance only degrades at a GEMM size two times larger than the baseline.

\begin{figure}[t]
    \centering
    \begin{subfigure}{\linewidth}
        \centering
        \includegraphics[width=0.95\linewidth]{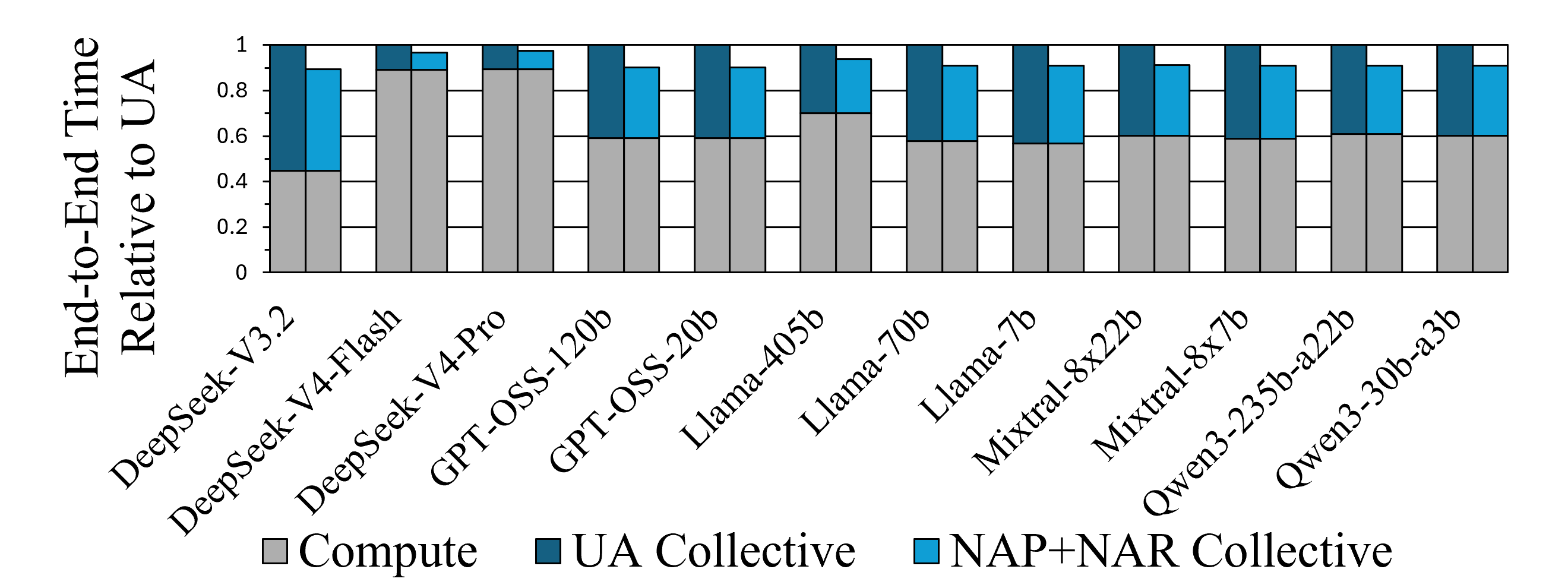}
        \vspace{-.4em} %
        \caption{Prefill. Lowest latency batch/parallelism per model.}
        \label{fig:e2e_prefill}
    \end{subfigure}

    \begin{subfigure}{\linewidth}
        \centering
        \includegraphics[width=0.95\linewidth]{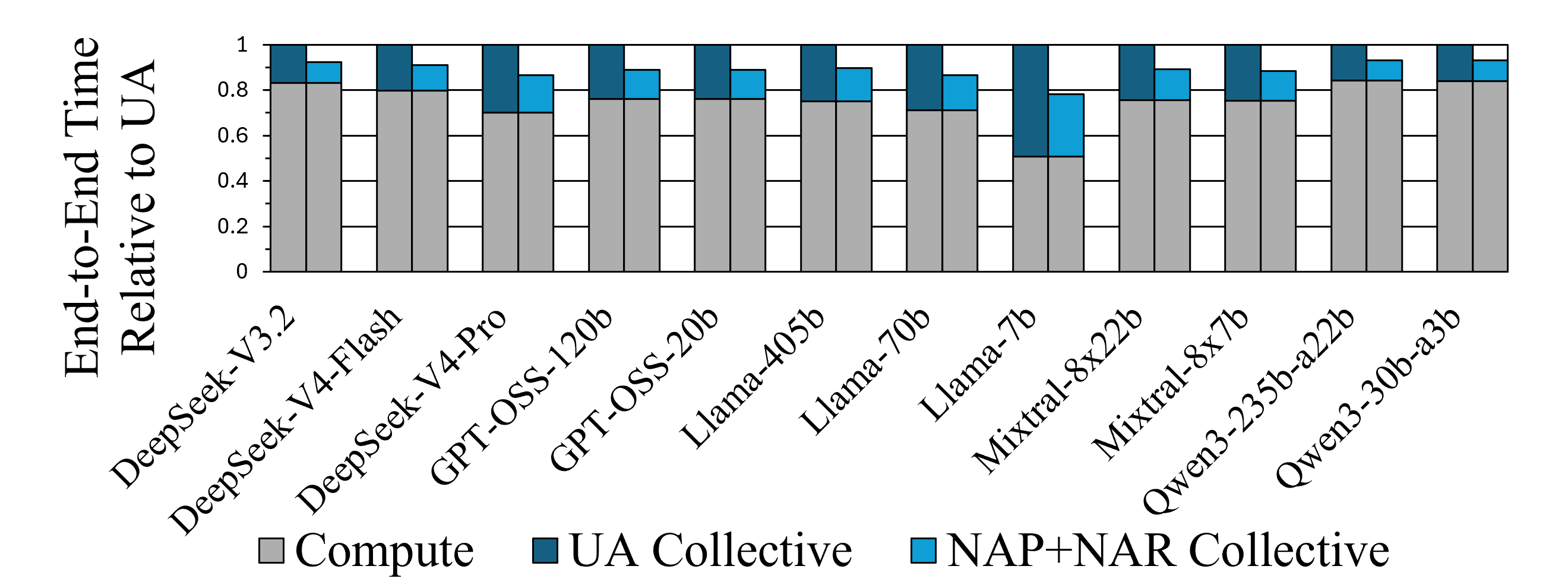}
        \vspace{-.4em} %
        \caption{Decode. Lowest latency batch/parallelism per model.}
        \label{fig:e2e_decode}
    \end{subfigure}

    \vspace{-.4em}

    \caption{End-to-end time (lower is better) of 64-GPU configuration relative to baseline. Compute (gray) includes compute and local memory operations. Baseline (UA, dark blue) and \ours{} (NAP+NAR, light blue) are collective communication.}

    \label{fig:e2e}
\end{figure}

\subsection{End-to-End LLM Execution}\label{nuna_e2e}

Finally, we evaluate the effect of \ours{}-aware techniques to the end-to-end scenario.
We apply the full~\nap{}+\nar{} scheme on 12 models across 2--64 GPUs.
\autoref{fig:e2e} specifically captures the decomposed runtime for 64-GPU results, where the communication is most exposed.

Averaged (\ie{} geometric mean) over all models and GPU counts, \ours{}-aware execution reduces prefill TTFT by 6\% (up to 11\%) and decode TPOT by 7\% (up to 28\%). These end-to-end reductions follow directly from faster collectives:~\nap{}+\nar{} improves communication time by a geomean of 1.32$\times$ for prefill collectives and 1.56$\times$ for decode collectives.
The larger decode benefit reflects its smaller, latency-bound messages.
We note Amdahl's law bounds the end-to-end gain by the compute-to-communication ratio.
Because compute is unchanged, exposed communication ratio is approximately 4:1 for prefill and 5:1 for decode, respectively.

In prefill (\autoref{fig:e2e_prefill}), collectives occupy a substantial share of
runtime but are comparatively large, so the per-collective (and end-to-end) gains are modest. In decode (\autoref{fig:e2e_decode}), collectives are small and latency-bound and see the largest reductions. As a representative point, the latency-optimal Llama-70B configuration (eight-way tensor parallelism with 64 GPUs) achieves 8\% and 16\% end-to-end speedups for prefill and decode, respectively.
\allreduce{} of partial activations recovers the most time, at a per-collective geomean of 1.54$\times$.
\alltoall{} expert token routing dominates MoE communication, which~\ours{} improves by 1.48$\times$ on average.
Consistent with our microbenchmarks, small messages get more benefit from \ours{}: collectives up to 1\,MB speed up 1.5–1.9$\times$, decaying to 1.1$\times$ by 64\,MB.

\section{Related Work}\label{sec:related}

\paratitle{Dynamic or Reactive}
Arunkumar et al.~\cite{arunkumar2017mcm} introduce NUMA-aware threadblock scheduling
with a first-touch physical page allocation policy.
However, the co-location policy neither explicitly controls the physical locations of the threadblocks on the compute die nor the physical locations of the memory chunks they access.
Therefore, they cannot optimize for \ours{}, as the \ours{} effect arises from the physical distances between CUs, memory, and I/O ports.
Milic et al.~\cite{milic2017beyond} propose dynamically changing link direction and caching policies to improve phased, asymmetric inter-GPU traffic.
This approach is inapplicable to collectives, which are low-latency and symmetric.
Reactive solutions may incur prohibitive penalties for page migrations of up to 20--50\,$\mu$s~\cite{zheng2016towards}, which can be avoided if the pages were statically allocated.

\paratitle{Static Analysis}
Kim et al.~\cite{kim2017coda} perform static analysis to co-locate threadblocks and memory chunks
to reduce off-chip traffic.
Several works introduce compiler and static profiling techniques for threadblock allocation to reduce inter-GPU communication~\cite{cabezas2015automatic} or improve cache locality~\cite{li2017locality,chen2017improving}.
Khairy et al.~\cite{khairy2020locality} propose locality-aware data management~(LADM) to manage memory and threadblock scheduling to optimize for multi-die GPUs.
LADM allocates compute and memory to reduce the amount of off-chip traffic.
However, reducing inter-GPU traffic by co-locating chunks and threadblocks is not applicable to collectives, as collective communications are inherently inter-GPU.
Furthermore, topology-specific information, critical for \ours{} mitigation, is under-specified and cannot be extended.
Kim et al.~\cite{kim2023locality} use static analysis to schedule threadblocks, which improves sharing, but the work is only applicable to convolutions.
Zhu et al.~\cite{zhu2024spgpu} propose spatially programmed GPU~(SPGPU) to allow programmer hints on threadblock and data placement.
However, it is based on tiled and strided access patterns, lacking generality, and appears to be a simplified version of others.

\section{Conclusions}\label{sec:conclusion}

Physically large, multi-die GPUs ensure their integration into future AI systems.
However, we should address the elongated wire delays to fully achieve their potential.
Prior works focused on compute-memory locality within a socket.
We emphasize that optimizing inter-GPU network operations, such as latency-sensitive collectives, requires the same attention to the increasing spatial effects.
In this paper, we introduce the concept of \ours{}.
We demonstrate how \ours{}-aware routing and \ours{}-aware placement techniques speed up collectives up to 1.8$\times$.
For latency-sensitive collectives running on next-generation GPU devices, we propose that hardware designers and programmers should consider \ours{} effects to maximize the performance of these impressive systems.

\section*{Acknowledgments}

We thank our colleagues Moumita Dey, Eris Furkan, Vinay Ramakrishnaiah, and Ruchi Shah for their revision advice and helpful discussions. We also thank Ganesh Dasika and Gabriel Loh for their reviews to improve the paper.

AMD, the AMD Arrow logo, AMD Infinity Fabric, AMD Instinct, and combinations thereof are trademarks of Advanced Micro Devices, Inc.
Other product names used in this publication are for identification purposes only and may be trademarks of their respective companies.

\bibliographystyle{ACM-Reference-Format}
\bibliography{refs}

\end{document}